# Tuning the Band Topology of GdSb by Epitaxial Strain

Hadass S. Inbar[1*&], Dai Q. Ho[2,3&], Shouvik Chatterjee[4#], Aaron N. Engel[1], Shoaib Khalid[2†], Connor P. Dempsey[4], Mihir Pendharkar[4♦], Yu Hao Chang[1], Shinichi Nishihaya[1‡], Alexei V. Fedorov[5], Donghui Lu[6], Makoto Hashimoto[6], Dan Read[4,7], Anderson Janotti[2], Christopher J. Palmstrøm[1,4*]

[1]*Materials Department, University of California Santa Barbara, Santa Barbara, CA 93106, USA*

[2]*Department of Materials Science and Engineering, University of Delaware, Newark, DE 19716, USA*

[3]*Faculty of Natural Sciences, Quy Nhon University, Quy Nhon 59000, Vietnam*

[4]*Electrical and Computer Engineering Department, University of California Santa Barbara, Santa Barbara, CA 93106, USA*

[5]*Advanced Light Source, Lawrence Berkeley National Laboratory, Berkeley, CA 94720, USA*

[6]*Stanford Synchrotron Radiation Lightsource, SLAC National Accelerator Laboratory, CA, USA*

[7]*School of Physics and Astronomy, Cardiff University, Cardiff CF24 3AA, UK*

Rare-earth monopnictide (RE-V) semimetal crystals subjected to hydrostatic pressure have shown interesting trends in magnetoresistance, magnetic ordering, and superconductivity, with theory predicting pressure-induced band inversion. Yet, thus far, there have been no direct experimental reports of interchanged band order in RE-Vs due to strain. This work studies the evolution of band topology in biaxially strained GdSb (001) epitaxial films using angle-resolved photoemission spectroscopy (ARPES) and density functional theory (DFT). We find that biaxial strain continuously tunes the electronic structure from topologically trivial to nontrivial, reducing the gap between the hole and the electron bands dispersing along the [001] direction. The conduction and valence band shifts seen in DFT and ARPES measurements are explained by a tight-binding model that accounts for the orbital symmetry of each band. Finally, we discuss the effect of biaxial strain on carrier compensation and magnetic ordering temperature.

---

[&]These authors contributed equally to this work.

Present Addresses: # Department of Condensed Matter Physics and Materials Science, Tata Institute of Fundamental Research, Mumbai 400005, India

† Department of Physics, School of Natural Sciences, National University of Science and Technology, Islamabad 44000, Pakistan

♦ Department of Materials Science and Engineering, Stanford University, Stanford, CA, 94305 USA

‡ Department of Physics, Tokyo Institute of Technology, Tokyo, 152-8551, Japan

*Author to whom correspondence should be addressed: hadass@ucsb.edu (H.S.I.), cjpalm@ucsb.edu (C.J.P.)

## I. INTRODUCTION

Strain engineering of low-dimensional topological quantum materials serves as a powerful approach to manipulating electronic band structures, thereby controlling topological phase transitions and transport behavior[1]. For example, strained HgTe quantum wells grown in the tensile and compressive regimes were shown to transition from a semimetallic to a two-dimensional topological insulator (TI) system, respectively[2]. Despite the promise of topological state tuning, strain studies of quantum materials as thin films are typically restricted to local, defect-induced strain gradients[3–5] or to strain levels below 1% strain[6,7] in the case of uniform strain in lattice-mismatched growths. In TIs such as the group V-chalcogenides ($X_2Z_3$, $X$=Bi, Sb; $Z$=Te, Se), unstrained growths occur even on substrates with high lattice mismatch due to the weak bonding between the van der Waals layers[8,9]. In addition to the challenge of stabilizing highly strained pseudomorphic topological materials, visualizing band structure modifications as a function of strain/pressure has been difficult in both bulk single crystals and thin films. In single crystals, large pressure cells are difficult to implement, and when using mechanical strain tuning apparatus special care is needed to ensure the application of uniform strain[10–13]. For thin films, there are limited reports combining strained film growth with direct spectroscopic tools such as angle-resolved photoemission spectroscopy (ARPES), with a few exceptions in oxide films[14,15].

Recent reports of bulk rare-earth monopnictide (RE-V) crystals under hydrostatic pressure reveal the emergence of a superconducting phase transition in nonmagnetic RE-Vs[16–18], and theoretical predictions suggest potential strain and pressure-induced transitions in band topology[19–23]. In addition to observing a strain-driven topological phase transition in the RE-V system, strain studies of RE-Vs are highly relevant for spintronic-based applications as another control knob to tune magnetoresistance and magnetic ordering in RE-V thin films. Finally, coupled with III-V semiconductors, RE-V thin films and particles have shown many potential device applications[24], including buried metallic contacts[25], THz emitters and detectors[26,27], thermoelectrics[28,29], plasmonic heterostructures[30], and diffusion barriers[31]. Therefore, straining RE-V thin films and thickness tuning present another avenue to control the functional properties of these magnetic semimetals, specifically by modifying magnetic exchange interactions and the charge carrier ratio in these otherwise electron-hole-compensated semimetal systems.

Here, we use ARPES to study the evolution of the electronic structure of GdSb thin films grown by molecular beam epitaxy (MBE) subjected to 2% tensile (+2%) and 2% compressive (-2%) biaxial strain. GdSb belongs to the RE-V family of compounds and is particularly interesting due to the relatively small electron-hole band energy gap that can be inverted via attainable strain/hydrostatic pressure, resulting in a nontrivial $\mathbb{Z}_2$ topological invariant classification[19]. We demonstrate the ability to tune the band gap and Néel temperature ($T_N$) in strained GdSb thin films and thereby control the topological phase transition from a trivial to a nontrivial state. GdSb thin films also present high magnetoresistance[32], have a type-II antiferromagnetic ordering at nearly the highest temperature of all RE-V ($T_N$ = 24 K)[33], and can be epitaxially integrated with III-V semiconductors[34], see Fig. 1(a-c). This approach to band engineering via epitaxial strain can be broadly applied to a wide range of RE-V antiferromagnet semimetals.

## II. MATERIALS AND METHODS

GdSb has a lattice parameter of $a$ = 6.219 Å between InSb (6.479 Å) and GaSb (6.096 Å)/AlSb (6.136 Å), allowing high tensile and compressive biaxial strain by varying the underlying semiconducting III-V buffer layer structure as shown in Fig. 1(d). MBE was used to grow epitaxial GdSb (001) thin films on $In_xGa_{1-x}Sb$/$In_xAl_{1-x}Sb$ buffer layers nucleated on a GaSb (001) substrate. For photoemission and scanning tunneling microscopy (STM) studies, p-type doped substrates and p-type $In_xGa_{1-x}Sb$ buffer layers were used. By changing the Ga/In or Al/In concentration in the buffer layer, the in-plane lattice parameter was adjusted before GdSb growth, as shown in Fig. 1(d). For magneto-transport measurements, undoped $In_xAl_{1-x}Sb$ buffer layers and epi-ready undoped GaSb (001) wafers (Wafer Technology Ltd.) were used. Further details on the GdSb growth window, ARPES measurement conditions, and electronic characterization of lattice-matched unstrained films are detailed in our previous report[32]. The growth of strained films was studied *in situ* with reflection high-energy electron diffraction, STM, and confirmed *ex situ* with x-ray diffraction reciprocal space map (RSM) measurements. Grazing incidence RSM of (226) reflections for 4-nm-thick GdSb films are shown in Fig. 1(b-c), confirming that the layers remain pseudomorphically strained to the buffer layer. *In situ* STM scans of the 2% compressive strained GdSb film grown directly on GaSb in Fig. 1(e) and on metamorphic III-V buffer layers[32] confirm the growth of a smooth and continuous GdSb film with terrace step heights consistent with half of a unit cell. We investigated the electronic structure of GdSb theoretically using density functional theory (DFT) and the screened hybrid functional of Heyd, Scuseria, and Ernzerhof (HSE06)[35,36] with 25% of exact exchange and accounting for spin-orbit coupling, as implemented in the VASP code[37,38]. See the supplementary material for additional details on MBE growth, ARPES measurements, and DFT calculations.

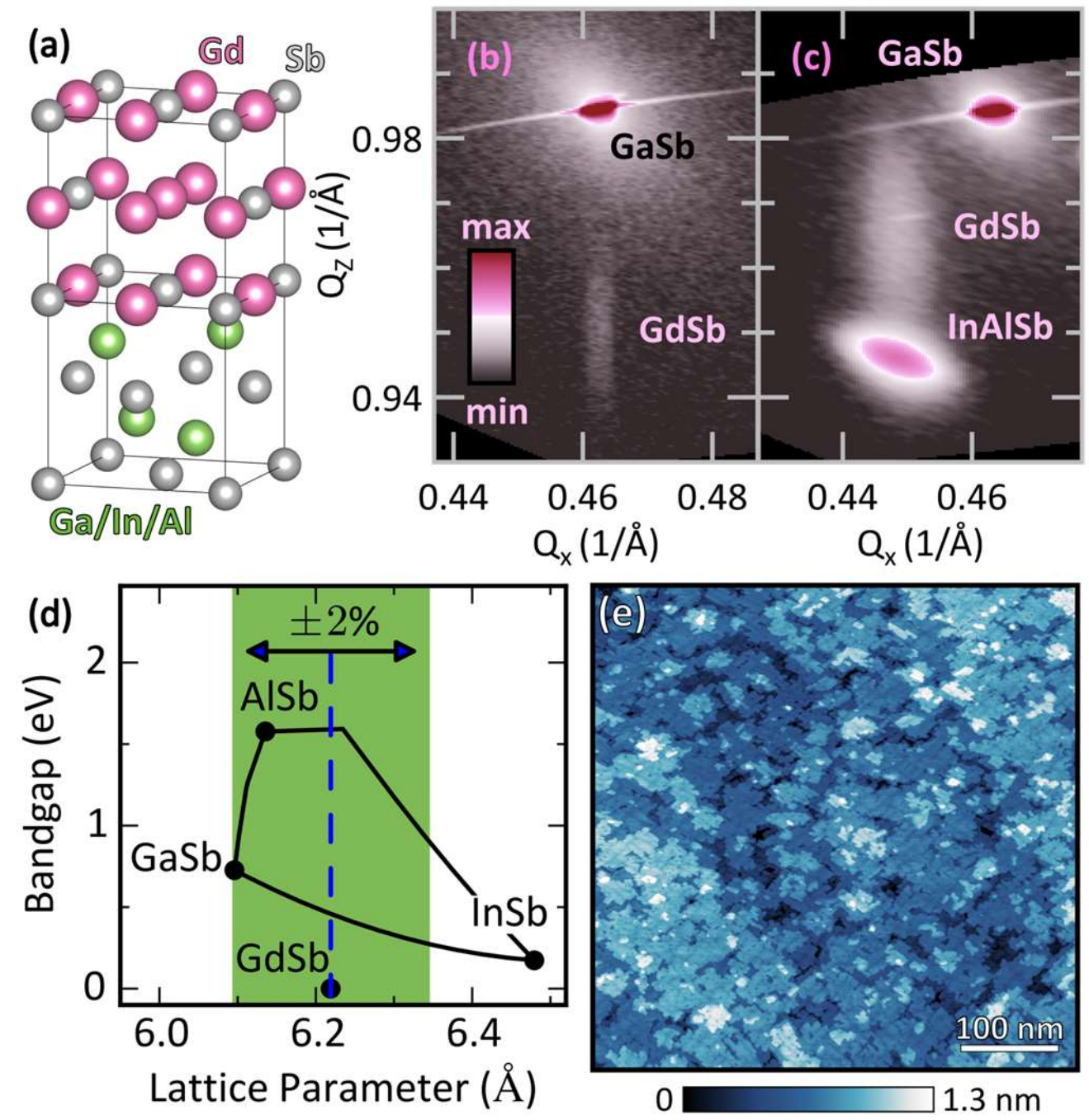

**Fig. 1.** (a) Crystal structure and epitaxial relationship of the rocksalt GdSb / zincblende III-V (001) orientation. RSM of the (226) reflection in (b) −2% and (c) +2% strained GdSb films, demonstrating coherent growth to the underlying III-V layer. (c) Biaxial strain window of GdSb and range of III-V band gaps and lattice parameters used for buffer layer growth. (d) STM image of the −2% strained film: 4 nm GdSb / GaSb (001) (V = −0.5 V, I=1 nA).

## III. RESULTS AND DISCUSSION

The Fermi surface of GdSb is composed of two hole pockets (β,δ) at the Brillouin zone center (Γ), a third spin-orbit split-off band (γ) positioned below the Fermi level, and three ellipsoidal electron pockets (α) at the Brillouin zone edge ($X_1$, $X_2$, $X_3$, the $X_3$ high-symmetry point transforming to the $Z$ point in the tetragonal *I4/mmm* space group under biaxial strain), see Fig. 2. In Figs. 2-4 we monitor the band topology evolution under strain, and address the effect of finite thickness quantization on the additional subbands observed in ARPES and modeled with DFT. DFT-calculated and ARPES-extracted Fermi wave vectors and band extrema positions for both strain values are summarized in Table 1.

Fig. 2(a-d) highlights the ARPES high-symmetry cuts studied for the electron pockets located at $X_{1,2}$ points in the film plane. Due to the high $k_z$ broadening expected for the vacuum ultraviolet light used in the ARPES measurements[39], the scans of the electron pocket in Fig. 2(d) present both the minor and major axes of the ellipsoidal electron pocket, the latter projecting from the neighboring Brillouin zone in Fig. 2(a-b). ARPES of the electron pockets at $X_{1,2}$ (Fig. 2(a-d)) shows an increase in the bandwidth and major axis Fermi wave vector upon compressive strain, with the band minima shifting from $\alpha_{X_{1,2}}^{+2\%} = -0.375$ eV to $\alpha_{X_{1,2}}^{-2\%} = -0.440$ eV. The hole band

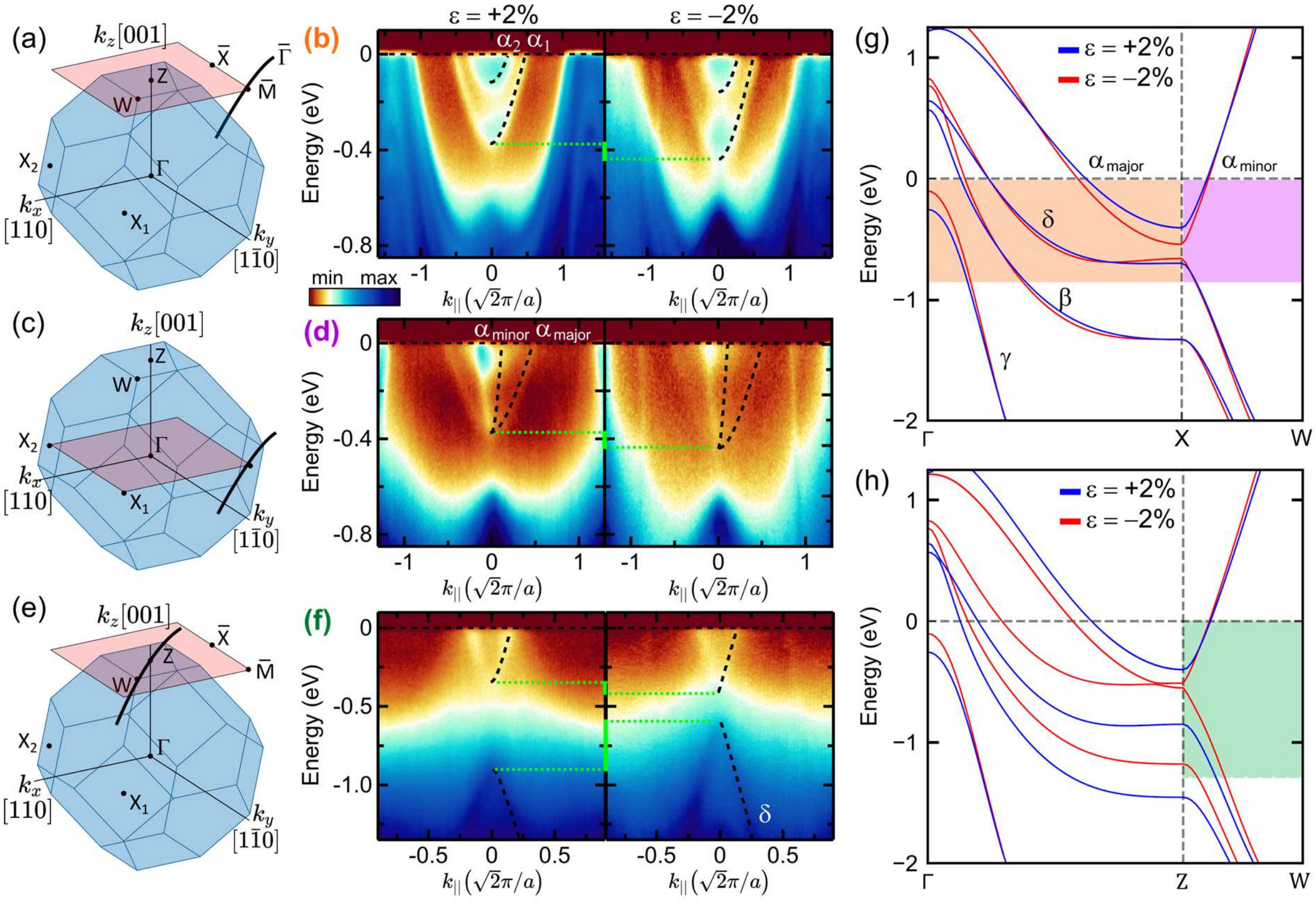

**Fig. 2.** Band dispersion of electron and hole bands in GdSb films studied with ARPES and strain-induced modifications for ε = +2% tensile (left) and ε = −2% compressive (right) biaxial strain. (a), (c), and (e) Schematics of the bulk Brillouin zone projected to the (001) surface Brillouin zone, showing the measured $k_z$ plane (pink square) and *E-k* spectra directions (black line). $\bar{\Gamma} - \bar{M} - \bar{\Gamma}$ cut along the in-plane electron pockets $X_{1,2}$ for (a,b) semimajor axis ($\Gamma - X_1 - \Gamma$) measured at $k_z$=$Z$ with a photon energy of 94 eV, and (c,d) semiminor axis ($W - X_2 - W$) measured at $k_z$= Γ with a photon energy of 60 eV. (e,f) $\bar{M} - \bar{\Gamma} - \bar{M}$ cuts of the out-of-plane electron pocket semiminor axis ($W - Z - W$) measured at $k_z$=$Z$ with a photon energy of 88 eV. Black dotted lines are hyperbolic fits to the band dispersions and the green dotted lines highlight the band shifts. See Table 1 for the Fermi wave vectors and band extrema extracted from the fits. (g-h) DFT-calculated band structures for ε = +2, −2% along (g) the in-plane high-symmetry points and (h) film plane normal direction. Fermi levels were set at 0. Shaded regions in (g-h) highlight the *E-k* cuts in panels (b,d,f).

extrema in the film plane remain largely unchanged: $\delta_{X_{1,2}}^{+2\%} = -0.66$ eV, $\beta_{X_{1,2}}^{+2\%} = -1.41$ eV, and $\delta_{X_{1,2}}^{-2\%} = -0.68$ eV, $\beta_{X_{1,2}}^{-2\%} = -1.45$ eV (see valence band pockets in Fig. 4 and summary of DFT and fit values in Table 1). These relatively small changes in the bands lying in the film-plane agree with our DFT calculations in Fig. 2(g) which predict only a small shift in the electron pockets at $X_{1,2}$ as a function of biaxial strain.

Fig. 2(e-f) shows the electron pocket at the $Z$ high-symmetry point, positioned along the film plane normal. Fig. 3 depicts the expanded momentum range of the $E$-$k$ cuts in Fig. 2(e-f), capturing electron and hole pockets lying perpendicular to the film plane ($Z$ for $k_{||} = 0$) and in the film plane ($X_{1,2}$ high-symmetry points from neighboring Brillouin zones projecting to $\bar{M}$ at $k_{||} = \frac{2\pi}{a}$). Fitting the electron and hole bands at $Z$ along $W - Z - W$ in (Fig. 2(f) and Fig. 3), we see a downward shift in the electron pocket from $\alpha_Z^{+2\%} = -0.34$ eV to $\alpha_Z^{-2\%} = -0.41$ eV, and for the valence bands an upward shift: $\delta_Z^{+2\%} = -0.9$ eV to $\delta_Z^{-2\%} = -0.59$ eV, and $\beta_Z^{+2\%} = -1.45$ eV to $\beta_Z^{-2\%} = -1.31$ eV. The influence of epitaxial strain on the hole bands, primarily at $Z$ (and only small shifts for the bands dispersing in-plane along $\Gamma - X - W$), agrees with our calculations in Fig. 2(g-h) and earlier DFT calculations performed for LaSb[40].

Our DFT calculations in Fig. 2(g-h) show that at $\varepsilon = -2\%$ GdSb transitions into a topological semimetal state as the hole and electron bands anti-cross along $\Gamma - Z$ and are inverted at $Z$. In contrast, the in-plane electron and hole bands at $X_{1,2}$ remain gapped. Due to high $k_z$ broadening in Fig. 2(e-f) and Fig. 3, the valence band pockets at $k_z=\Gamma$ also project to the $k_z=Z$ plane leading to a blurred background intensity preventing the observation of the expected topological surface states (TSS) for the compressively strained film (TSS in RE-Vs typically have a weaker spectral intensity compared to the bulk bands[41,42]), and leading to a larger error bar in our estimation of electron band minima at *Z*. Nevertheless, the ARPES dispersions for the hole bands at *X* and *Z*, and the electron bands lying in the film plane are consistent with our DFT calculations and support the predicted band gap reduction scenario at *Z* moving from tensile to compressive strain.

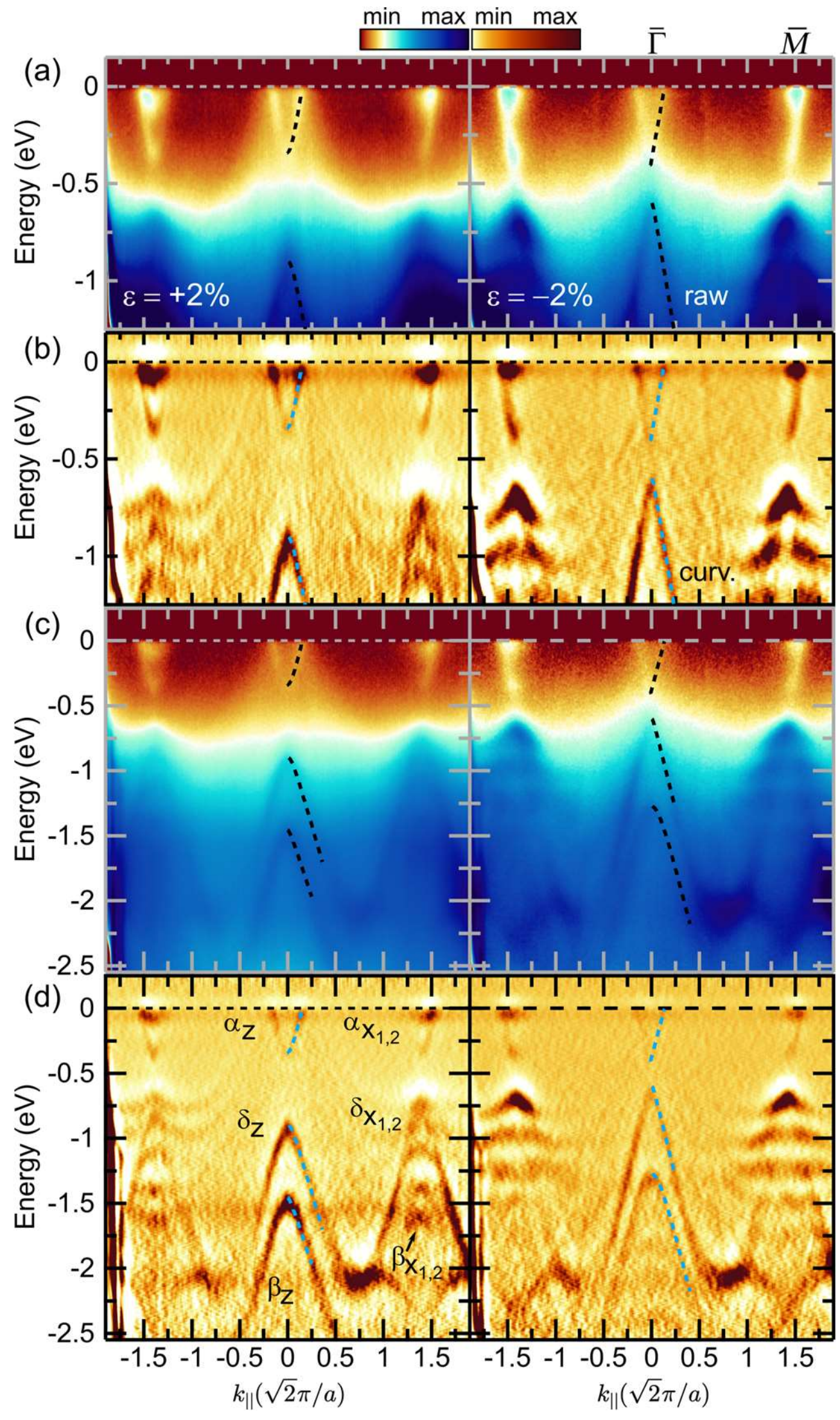

**Fig. 3.** $E$-$k$ dispersion along $\bar{M} - \bar{\Gamma} - \bar{M}$ ($X_{1,2} - W - Z - W - X_{1,2}$) in biaxial strained GdSb films measured with a photon energy of 88 eV, an expanded range of the cuts shown in Fig. 2(e-f). (a,c) Raw data and (b,d) curvature plots of the raw data presenting the band dispersions (a,b) near the Fermi level, and (c,d) over a wider energy range including the $\beta$ hole pocket maximum.

We have further checked the predicted topological nature of the strained GdSb by evaluating the $\mathbb{Z}_2$ strong topological index $\nu_0$ according to the band parity product criteria[43] considered at eight time-reversal inversion momenta (TRIM) points: *Γ*, 4 *L*, 2 *X*, and *Z* where $(-1)^{\nu_0} = \prod_{i=1}^{8} \delta_i$ , $\delta_i$ being the parity product at each TRIM point for all occupied bands. Time-reversal symmetry ($\Theta$) and primitive-lattice translation symmetry ($T_{1/2}$) in GdSb are broken, however their combination is preserved (S=$\Theta T_{1/2}$), enabling the classification of the topological nature using the $\mathbb{Z}_2$ topological invariant[44]. In the unstrained and tensile cases, the $\mathbb{Z}_2$ invariant $\nu_0 = 0$, demonstrating a trivial topological state. In contrast, we observe a change in the parity product at the *Z* point (from $+$ to $-$) in the compressively strained case, resulting in a $\mathbb{Z}_2$ index $\nu_0 = 1$ which indicates a nontrivial topological band structure. Thus, the 2% compressive GdSb appears to lie within the transition region between a strong topological insulator with nonzero weak indices and a $\mathbb{Z}_2$trivial topological state.

**Table 1.** Fermi surface of 4-nm-thick strained GdSb (001) films. Band maximum/minimum energy positions, Fermi wave vectors ($k_F$) for all bands, and the carrier density ratio, obtained from ARPES measurements and DFT calculations. Band extrema are reported for all quantum well subbands observed via ARPES.

| | | | +2% (tensile strain) | | −2% (compressive strain) | |
|---|---|---|---|---|---|---|
| | | | ARPES | DFT | ARPES | DFT |
| α $k_F$ (Å$^{-1}$) | Minor $W-X$ | $X_{1,2}$ | 0.085 (±0.02) | 0.103 | 0.084 (±0.02) | 0.110 |
| | | Z | 0.11 (±0.05) | 0.108 | 0.089 (±0.02) | 0.103 |
| | Major $\Gamma-X$ | $X_{1,2}$ | 0.356 (±0.02) | 0.374 | 0.371 (±0.03) | 0.423 |
| | | Z | NA | 0.364 | NA | 0.434 |
| α Band Extrema (eV) | | $X_{1,2}$ | $\alpha_1$: −0.375 (±0.004)<br>$\alpha_2$: −0.118 (±0.006) | −0.405 | $\alpha_1$: −0.440 (±0.004)<br>$\alpha_2$: −0.157 (±0.004) | −0.540 |
| | | Z | -0.34 (±0.02) | −0.398 | −0.41 (±0.02) | −0.512 |
| δ $k_F$ (Å$^{-1}$) | $\bar{M}-\bar{\Gamma}-\bar{M}$ | | 0.230 (±0.030) | 0.238 | 0.184 (±0.030) | 0.248 |
| | $\bar{X}-\bar{\Gamma}-\bar{X}$ | | 0.123 (±0.01) | 0.175 | 0.124 (±0.018) | 0.189 |
| β $k_F$ (Å$^{-1}$) | $\bar{M}-\bar{\Gamma}-\bar{M}$ | | 0.116 (±0.017) | 0.127 | 0.105 (±0.005) | 0.151 |
| | $\bar{X}-\bar{\Gamma}-\bar{X}$ | | 0.064 (±0.008) | 0.127 | 0.077 (±0.010) | 0.151 |
| γ Band Extrema (eV) | | Γ | −0.32 (±0.01) | −0.255 | −0.30 (±0.01) | −0.102 |
| | | $X_{1,2}$ | −3.205 (±0.05) | −3.33 | −3.12 (±0.05) | −3.27 |
| δ Band Extrema (eV) | | $X_{1,2}$ | $\delta_1$: −0.66 (±0.02)<br>$\delta_2$: −0.82 (±0.02)<br>$\delta_3$: −1.09 (±0.02) | -0.70 | $\delta_1$: −0.68 (±0.02)<br>$\delta_2$: −0.88 (±0.05)<br>$\delta_3$: −1.21 (±0.02) | −0.66 |
| | | Z | $\delta_1$: −0.90 (±0.02) | −0.855 | $\delta_1$: −0.59 (±0.08) | −0.55 |
| β Band Extrema (eV) | | $X_{1,2}$ | $\beta_1$: −1.41 (±0.01)<br>$\beta_2$: −1.62 (±0.03) | −1.33 | $\beta_1$: −1.45 (±0.03)<br>$\beta_2$: −1.70 (±0.05) | −1.33 |
| | | Z | $\beta_1$: −1.45 (±0.02) | −1.45 | $\beta_1$: −1.28 (±0.02) | −1.18 |
| $n_e/n_h$ | | | 1.52 | 1.09 | 1.85 | 1.11 |

In RE-Vs under hydrostatic pressure[19,21] all three $\Gamma - X$ high-symmetry directions are equivalent, yet epitaxial strain primarily affects bands in the direction normal to the film plane, i.e. along [001] ($\Gamma - Z$). The band gap changes in $\Gamma - Z$ can be explained using a simple tight-binding (TB) model, accounting for the orbital composition of the electron and hole bands near the Fermi level and the scaling of nearest-neighbor and next-nearest neighbor interactions with strain (see Fig. 5 and the supplementary material for details on the TB model). Based on the orbital-resolved DFT electronic band structure in Fig. 5(a), we construct a TB model that reproduces well the DFT-calculated band structure and the effect of strain on hopping terms (Table S1). The band structure of GdSb resulting from our TB parametrization is presented in Fig. 5(d) and Fig. S1 over a narrow and wide energy range, respectively. From Fig. 5(a), it is apparent that the out-of-plane electron pocket centered at $Z$ is mainly composed of Gd $d_{xy}$ orbitals, which form $dd\sigma$-like bonds in the [110] direction and $dd\pi$-like bonds along the [101] direction (highlighted in Fig. 5(b), with $dd\delta$ hopping being negligible, i.e., close to 0). The heavy- (δ) and light- (β) hole bands consist of Sb $p_x + p_y$ orbitals along $\Gamma - Z$, forming three different hopping terms $t_{1,2} = pp\pi \pm pp\sigma$ $t_3 = pp\pi$. The split-off valence band (γ) is made up of $p_z$ orbitals. Moreover, $p$-$d$ mixing in GdSb through $pd\sigma$ and $pd\pi$ bond formation is necessary to describe the sharp conduction and valence band dispersions along $Z - W$. Similarly, the in-plane electron pockets at $X_1/X_2$ and δ hole band dispersions along the $\Gamma - X_{1,2}$ axis are composed of Gd $d_{yz}$ / $d_{xz}$ orbitals and Sb $p_y + p_z$ / $p_x + p_z$ orbitals, respectively.

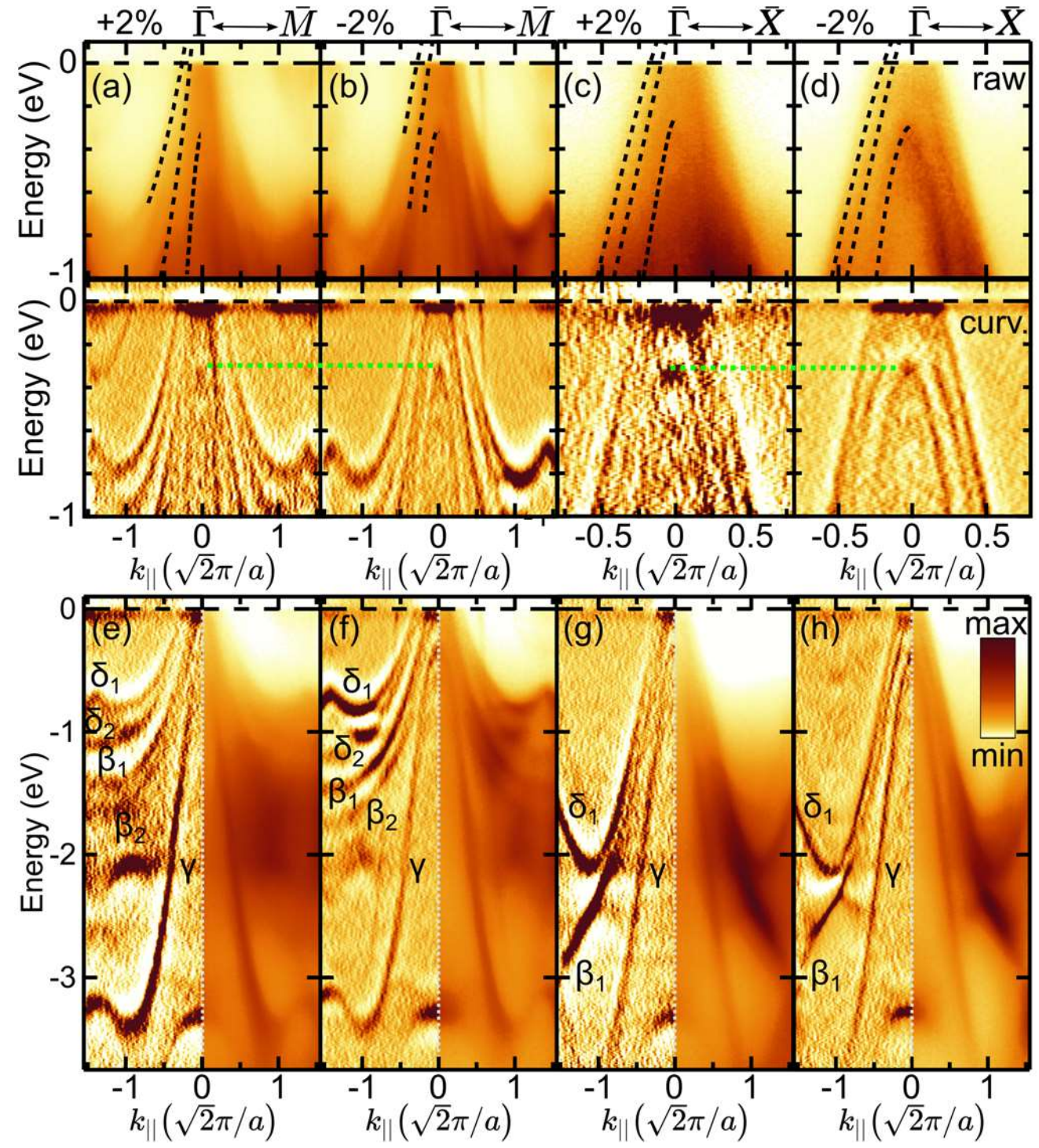

**Fig. 4.** *E-k* dispersion of hole pockets in biaxial strained GdSb films measured at $k_z = \Gamma$ (photon energy of 60 eV). (a-d) ARPES spectra near the Fermi level. The green dotted line highlighting the same γ valence band maximum position and the black lines show overlaid fits along (a,b) $\bar{M} - \bar{\Gamma} - \bar{M}$ and (c,d) $\bar{X} - \bar{\Gamma} - \bar{X}$. Top panel: raw data, bottom panel: curvature plots of the raw data. (e-h) Wider energy range of the same cuts in (a-d), showing the quantum well states. The plots on the right side present the raw data, and the plots on the left side display the curvature plot. Fermi wave vectors extracted from the fits to the valence band and band extrema in panels (e-h) are detailed in Table 1.

The DFT-calculated gap at $Z$, $E_g(Z)$, between the conduction and valence bands as a function of strain level is shown in Fig. 5(c). Compressive strain widens the bandwidth along $X - W$ and $Z - W$ for the hole and electron pockets (see Table 1 and Fig. 2(g-h)). Upon applying compressive strain, the orbital overlap increases between the in-plane hopping terms of the $p_x$ $p_y$ $d_{xy}$ orbitals, leading to increased dispersion in both the valence and the conduction bands. However, because the $d$ orbital hopping terms have a stronger distance dependence than the $p$ orbitals[45], shown in Fig. 5(c), the electron pocket has a more significant increase in its bandwidth moving from tensile to compressive strain. This behavior also explains the topological phase transition trend observed for RE-Vs due to lanthanide contraction[19,22]. Lighter lanthanide elements have both larger ionic radius[46] and larger unit cells[47], yet overall the ratio of the lanthanide ionic radius to the RE-V unit cell increases for the lighter lanthanides, leading to higher *d-d* orbital overlap eventually resulting in band inversion, despite the decrease in *p-p* orbital overlap.

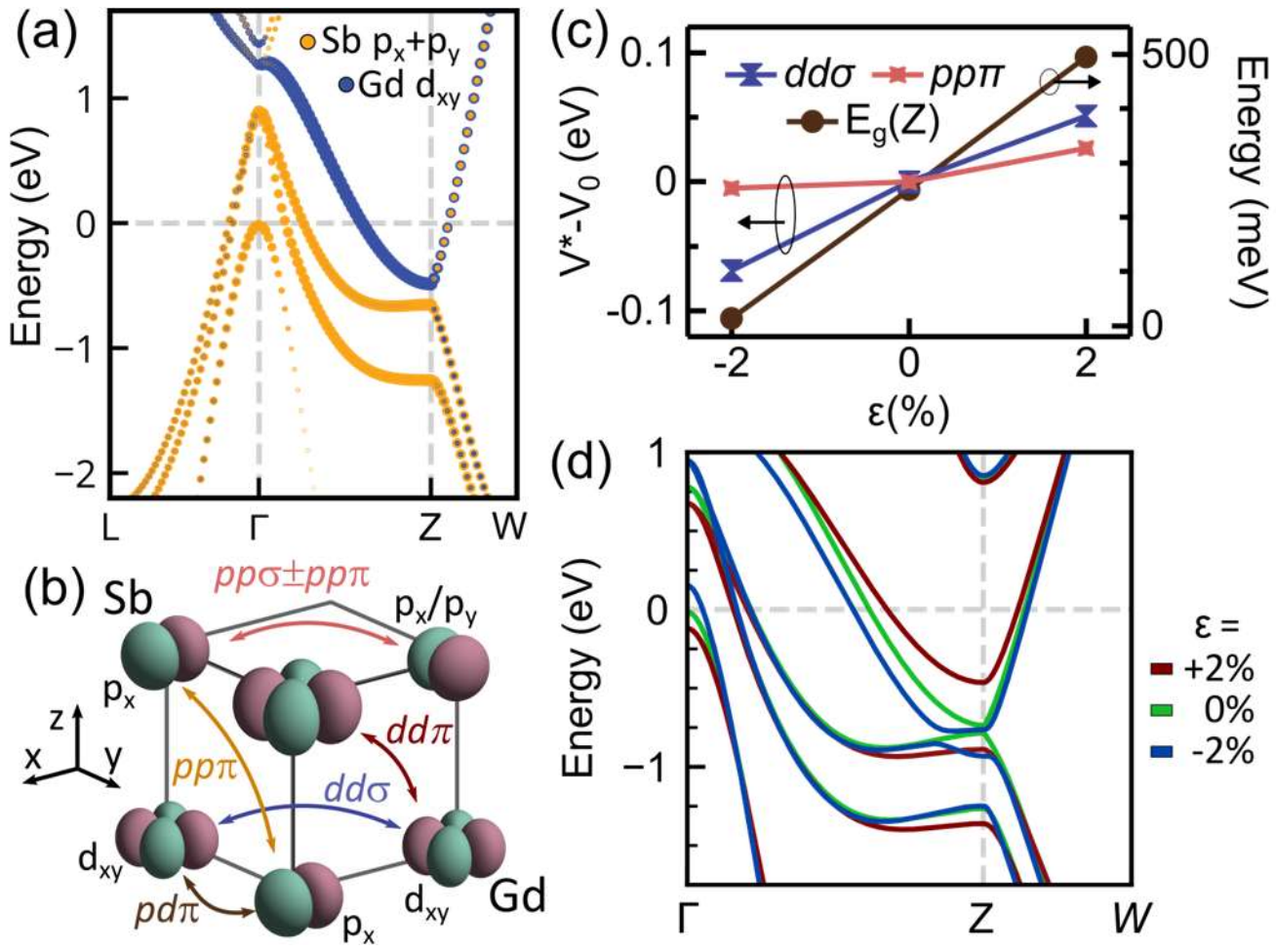

**Fig. 5.** Strain effect on orbital overlap in GdSb. (a) Nonmagnetic DFT calculations of the electronic band structure in GdSb and the orbital character of the DFT wavefunctions. (b) Illustration of relevant atomic orbitals and primary interaction paths at the *Z* point. (c) Tight binding hopping term decay rate with increasing strain (see supplementary material for details) and the evolution of the calculated gap at the TRIM *Z* point vs strain. (d) Tight binding band structure of GdSb and its dependence on strain level.

Strain-induced band inversion along $\Gamma - Z$ in Fig. 2(h) and Fig. 3 is reproduced in the TB model in Fig. 5(d) and explains the more substantial modifications observed in ARPES and measured in DFT for band dispersions composed of atomic orbitals distributed within the film plane. For the in-plane electron pockets at $X_{1,2}$ under compressive strain, the *d-d* orbital overlap in the (011)/(101) faces is affected by both the reduced distance along [010]/[100] and the slightly expanded out-of-plane lattice parameter along [001]. However, due to the small Poisson ratio of GdSb, the total distance between the $d_{xz}$/$d_{yz}$ orbitals decreases, leading to a slight reduction of the gap at the $X_{1,2}$. In conclusion, the TB model demonstrates the importance of both Gd-Gd *dd*σ and Sb-Sb *pp*π bonding in determining the degree of band inversion at the *Z* high-symmetry point.

Next, we map the quantum well states in the conduction and valence bands. Two electron subband pockets are present at both strain levels in Fig. 2(b). Scans of the hole pockets along $\bar{M} - \bar{\Gamma} - \bar{M}$ ($X_{1,2} - \Gamma - X_{1,2}$ in the bulk Brillouin zone) and $\bar{X} - \bar{\Gamma} - \bar{X}$ ($K - \Gamma - K$ in the bulk Brillouin zone) in Fig. 4 show multiple quantum well states and agree with the number of subbands seen in our DFT calculations in Fig. S2 for films of the same thickness. The same number of quantum well subbands and similar energy splitting for the biaxially strained films confirm the growth of atomically uniform films of the same thickness and comparable interface potentials when grown on GaSb (ε = −2%) and $In_{0.65}Ga_{0.35}Sb$ (ε = +2%) buffer layers.

Due to quantum size effects, the band extrema positions in the 4-nm-thick films (detailed in Table 1) are expected to be shifted to higher (lower) binding energies for the in-plane dispersing hole (electron) pockets at $X_{1,2}$ compared to the DFT calculations in Fig. 2(g) performed for bulk-like GdSb. DFT calculations in Figure S1 (see supplementary material), modeling the effect of quantum confinement in 13 ML of GdSb (001) slabs, show that the electron pockets in the film plane at $X_{1,2}$ are experiencing quantum confinement, in contrast to the electron bands lying at *Z* which follow the bulk band-structure calculations. Therefore, pockets in the direction normal to the film plane are less susceptible to quantum confinement in the (001) plane due to their in-plane orbital composition.

A modest change in the concentration of all charge carriers is seen as a function of strain (see our earlier work[32] for details on the carrier density analysis). Overall, the charge carrier ratio increases with compressive strain from $\left(\frac{n_e}{n_h}\right)_{+2\%} = 1.52$ to $\left(\frac{n_e}{n_h}\right)_{-2\%} = 1.85$, suggesting that biaxial strain could serve as another degree of freedom to tune magnetoresistance in RE-Vs. As in past observations for LuSb[48] (a nonmagnetic RE-V analog), due to quantum confinement effects in the 4-nm-thick films, the Fermi surface area of the hole pockets and the electron pockets in the strained thin films is slightly smaller than the values extracted via ARPES for thicker unstrained GdSb films[32]. The electron-rich carrier ratio measured for both thin films, deviating from exact compensation in unstrained bulk GdSb, agrees with our earlier studies of quantum confinement effects in RE-Vs[48].

Finally, we address the effect of strain on the magnetic properties of GdSb. Due to the absence of orbital angular momentum in the $4f^7$ configuration of the $Gd^{3+}$ ion, GdSb represents an ideal isotropic Heisenberg model system for studying magnetic exchange interactions. GdSb can be considered a parent compound of the half-Heusler structure GdPtV (V = Bi, Sb), which shows complex behavior such as an antiferromagnetic to ferromagnetic transition in GdPtSb due to strain gradients[5] and chiral anomaly and anisotropic magnetotransport in the predicted Weyl semimetal GdPtBi[49]. A 2.6 K increase in the Néel temperature ($T_N$) from +2% tensile (24.3 K) to −2% compressive (26.9 K) strained films is observed in Fig. S3. Based on our TB model, a reduction in the lattice parameter results in increased *p-d* orbital hopping, which in turn leads to a higher $T_N$ (see further details in the supplementary material). To our knowledge, this is the first study showing the direct impact of epitaxial strain on superexchange *p-d* hopping in a RE-V and could guide future efforts in strain tuning the magnetic ordering temperature of other materials beyond the RE-V family. Building on these results, the effect of strain/pressure in other RE-Vs with more complex magnetic behavior, such as Ce-V[50] and Eu-VI[51], can be modeled or applied to semiconducting RE-V nitrides such as ScN[52] and GdN[53].

## IV. CONCLUSIONS

In summary, we have followed with ARPES and DFT the evolution of the bulk band structure in biaxial strained GdSb quantum wells and demonstrated the tuning of band gaps in RE-Vs through epitaxial strain. We report the successful growth of strained GdSb films integrated with a conventional III-V semiconducting substrate, and the resulting trends in magnetic ordering temperature and charge carrier ratios are discussed. The synthesis of high-quality epitaxial GdSb is an important step toward practical control of transport characteristics in magnetic Weyl semimetals. Our TB model based on nearest and next-nearest neighbor interactions describes well the electronic structure of GdSb. We have shown that biaxial compressive strain is expected to promote *d-d* hopping in the rare earth $t_{2g}$ conduction bands to a larger extent than the pnictogen *p* band hopping, resulting in band inversion and a higher electron carrier density. This work opens the door to future studies of strain-controlled topological phase transitions and semimetal-semiconductor transitions in RE-Vs[54] and RE-V derived compounds such as topological half-Heusler alloys (RE Pt/Pd V)[55–57].

## SUPPLEMENTARY MATERIAL

Additional details on ARPES surface preparation, DFT calculations, ARPES- and DFT-extracted Fermi wave vectors and band extrema, confinement effects, tight binding model construction, and magnetic properties of strained GdSb films.

## ACKNOWLEDGMENTS

Synthesis of thin films, development of a prototype ultrahigh vacuum suitcase, ARPES experiments, and theoretical work were supported by the U.S. Department of Energy (contract no. DE-SC0014388). Development of the growth facilities and low-temperature magnetotransport measurements were supported by the Office of Naval Research through the Vannevar Bush Faculty Fellowship under award no. N00014-15-1-2845. Scanning probe studies were supported by NSF (award number DMR-1507875). This research used resources of the Advanced Light Source (beamline 10.0.1.2), which is a DOE Office of Science User Facility under contract no. DE-AC02-05CH11231. Use of the Stanford Synchrotron Radiation Lightsource, SLAC National Accelerator Laboratory (beamline 5-2), is supported by the U.S. Department of Energy, Office of Science, Office of Basic Energy Sciences under Contract No. DE-AC02-76SF00515.We acknowledge the use of shared facilities of the NSF Materials Research Science and Engineering Center (MRSEC) at the University of California Santa Barbara (DMR 1720256). DFT calculations used the National Energy Research Scientific Computing Center (NERSC), a U.S. Department of Energy Office of Science User Facility operated under contract no. DE-AC02-05CH11231. H. S. I. gratefully acknowledges support from the UC Santa Barbara NSF Quantum Foundry funded via the Q-AMASE-i program under award DMR-1906325 and support for further developments of the vacuum suitcases. D.Q.H. acknowledges support from NSF through the University of Delaware Materials Research Science and Engineering Center, DMR-2011824.

## AUTHOR CONTRIBUTIONS

H.S.I. and C.J.P. conceived the study. Thin film growth, x-ray diffraction, and transport measurements were performed by H.S.I. with assistance from S.C., C.P.D., and M.P. ARPES measurements were performed by H.S.I. with assistance from S.C., A.N.E., Y.C. S.N., D.R., A.V.F., D.L., M.H. and analyzed by H.S.I. DFT calculations were performed by D.Q.H. with assistance from S.K., and A.J. supervising. H.S.I. and D.Q.H. constructed the tight-binding model. S.C., A.N.E., C.P.D., M.P. and designed ultra-high vacuum components and sample holders. The manuscript was prepared by H.S.I., D.Q.H., A.J., and C.J.P. C.J.P. supervised the experimental work. All authors discussed the results and commented on the manuscript.

# Supplemental materials for "Tuning the Band Topology of GdSb by Epitaxial Strain"

Hadass S. Inbar[1*&], Dai Q. Ho[2,3&], Shouvik Chatterjee[4#], Aaron N. Engel[1], Shoaib Khalid[2†], Connor P. Dempsey[4], Mihir Pendharkar[4♦], Yu Hao Chang[1], Shinichi Nishihaya[1‡], Alexei V. Fedorov[5], Donghui Lu[6], Makoto Hashimoto[6], Dan Read[4,7], Anderson Janotti[2], Christopher J. Palmstrøm[1,4*]

[1]*Materials Department, University of California Santa Barbara, Santa Barbara, CA 93106, USA*

[2]*Department of Materials Science and Engineering, University of Delaware, Newark, DE 19716, USA*

[3]*Faculty of Natural Sciences, Quy Nhon University, Quy Nhon 59000, Vietnam*

[4]*Electrical and Computer Engineering Department, University of California Santa Barbara, Santa Barbara, CA 93106, USA*

[5]*Advanced Light Source, Lawrence Berkeley National Laboratory, Berkeley, CA 94720, USA*

[6] *Stanford Synchrotron Radiation Lightsource, SLAC National Accelerator Laboratory, CA, USA*

[7] *School of Physics and Astronomy, Cardiff University, Cardiff CF24 3AA, UK*

## I. ARPES surface preparation and MBE growth

To reduce thin-film contributions to the electronic band structure, such as substrate charge transfer and quantum confinement[1], highly strained films as thick as possible (i.e., near the critical relaxation thickness) are studied. In GdSb, strong film-substrate interactions and interlayer bonding allow pseudomorphic growth at high strain levels. The critical thickness ($h_c$) is determined empirically as the onset of partial relaxation observed with reciprocal space mapping, with $h_c = 5.5\,nm$ in 2% compressively strained films. Consequently, all strained growths are limited to 4 nm in thickness. From the RSM-extracted in-plane and out-of-plane GdSb lattice parameters, we calculated the planar ($\varepsilon_{\parallel}$) and vertical ($\varepsilon_{\perp}$) strains and Poisson's ratio: $\frac{\varepsilon_{\perp}}{\varepsilon_{\parallel}} = \frac{-2v}{1-v}$ [2] such that $\nu_{exp} = 0.12 \pm 0.03$. The experimental Poisson's ratio agrees with our DFT-calculated value $\nu_{DFT} = 0.10$ and is comparable to earlier predictions made for other RE-Vs[3,4].

ARPES measurements at the Advanced Light Source (ALS) at beamline 10.0.1.2 were conducted at 11 K and at the Stanford Synchrotron Radiation Lightsource (SSRL) at beamline 5-2 at 20 K. Both were acquired with a Scienta DA30L hemispherical analyzer. ALS measurements were conducted for *in vacuo* transferred samples, where a custom-built vacuum suitcase with a base pressure <$10^{-10}$ Torr was used to transfer films from the growth chamber at UC, Santa Barbara, to beamline 10.0.1.2 at the ALS in Berkeley. The SSRL measurements at beamline 5−2 were performed for *ex situ* transferred films. To prevent oxidation of the GdSb films during the *ex situ* transfer to SSRL, a thick ~800 nm antimony capping layer was deposited with $Sb_2$ flux after GdSb growth during sample cooling down beginning at 230-130 °C. Before ARPES measurements, samples were heated to 420 ± 20 °C (calibrated by pyrometry) in an ultra-high vacuum chamber and held at that temperature for ~30 min to fully desorb the Sb cap. The potential thermal decomposition of the $In_xGa_{1-x}Sb$ buffer film limits the maximum annealing temperature. During the final stage of Sb desorption, a spike in pressure to 1e-8 Torr was observed, and the film surface visually transitioned from a shiny to a hazy-matt finish and back to a polished appearance. The thermal desorption window of the Sb cap was confirmed by scanning tunneling microscopy at UCSB and at the SSRL beamline by examination of the Sb 4*d* and Gd *f* core levels. All films showed no evidence of oxidation in XPS scans: a single binding energy component for the Gd 4*f* level was seen, and no oxygen 2*s* core levels were present. An Sb-related surface state was observed for all films and was remarkably stable for the compressive film, see sharp linear dispersions crossing the Fermi level at $k_F \sim 0.9$ Å$^{-1}$ in Fig. 2(d). The Fermi surface of the surface state differs from the expected electronic band structure of elemental Sb and might have originated from a stabilized square-net Sb-rich surface reconstruction[5].

## II. Tight binding model construction: hopping parameters and Hamiltonian construction

Our orbital composition determination in Fig. 5 shows the *p*- and *d*-orbital composition of the valence and conduction band, respectively, and agrees with ARPES measurements by Nummy *et al.*[16] assigning the orbital characters in the analogous La-Vs. In addition, we do not observe significant mixing between *s* orbitals and the group of *p* and *d* orbitals near the Fermi level. The TB Hamiltonian is thus constructed by 16x16 matrix elements, consisting of 8 atomic orbitals (Sb: $p_x$, $p_y$, $p_z$; and Gd: $d_{yz}$, $d_{zx}$, $d_{xy}$, $d_{x2-y2}$, $d_{z2}$) and accounting for spin-orbit coupling (SOC) which contains the atomic SOC parameter $\lambda_{Gd}$ and $\lambda_{Sb}$. These parameters are matched to the Gd *5d* and Sb *5p* atomic values, which are $\lambda_{Gd}$ = 0.43 eV and $\lambda_{Sb}$= 0.53 eV[22]. On-site energies are derived from the unstrained GdSb calculation: $Sb_p$= 4.190 eV, $Gd_d(e_g)$= 9.283 eV, and $Gd_d(t_{2g})$= 7.562 eV.

Taking the two-center approximation and considering only the nearest-neighbor Gd-Sb interactions and the next nearest-neighbor Gd-Gd and Sb-Sb interatomic couplings, the hopping terms are expressed by [23]:

$$t(p_x,p_y)_{[110]}=1/2\,(p_{Sb}p_{Sb}\sigma - p_{Sb}p_{Sb}\pi)$$
$$t(p_x,p_x)_{[110]}=1/2\,(p_{Sb}p_{Sb}\sigma + p_{Sb}p_{Sb}\pi)$$
$$t(p_{x/y},d_{xy})_{[100]/[010]} = p_{Sb}d_{Gd}\pi$$
$$t(d_{xy},d_{xy})_{[110]}=1/4(3d_{Gd}d_{Gd}\sigma + d_{Gd}d_{Gd}\delta)$$
$$t(d_{xy},d_{xy})_{[011]}=1/2(d_{Gd}d_{Gd}\pi + d_{Gd}d_{Gd}\delta)$$
$$t(d_{xy},d_{yz})_{[101]}=1/2(d_{Gd}d_{Gd}\pi - d_{Gd}d_{Gd}\delta)$$
$$t(p_{x/y},d_{x^2-y^2})_{[100]/[010]} = \frac{\sqrt{3}}{2}p_{Sb}d_{Gd}\sigma$$

The obtained TB parameters in the standard Slater-Koster notation are tabulated in Table S1. The Fermi level was set to 5.6 eV so that the DFT and TB dispersions match in position for unstrained GdSb.

Using the parameters listed in Table S1, the TB model is constructed using the open-source package Chinook[24]. The energy dispersions obtained from the TB model in Fig. 5(d) and Figure S1 reproduce well the ARPES and DFT results. Being isotropic, the TB model accounts for the strain-induced changes in hopping terms by using Slater-Koster parameters derived primarily from hopping terms in the (001) plane (except $t(d_{xy},d_{xy})_{[011]}$ and $t(d_{xy},d_{yz})_{[101]}$ in order to extract $d_{Gd}d_{Gd}\pi$ and $d_{Gd}d_{Gd}\delta$). The hopping term decay rate, $V^* - V_0$, is calculated based on the absolute difference between the strained ($V^*$) and unstrained ($V_0$) hopping terms. For example, $V(p_{Sb}p_{Sb}\pi)^{+2\%} - V(p_{Sb}p_{Sb}\pi)^{0\%} = -0.064 - (-0.090) = 0.026$ .

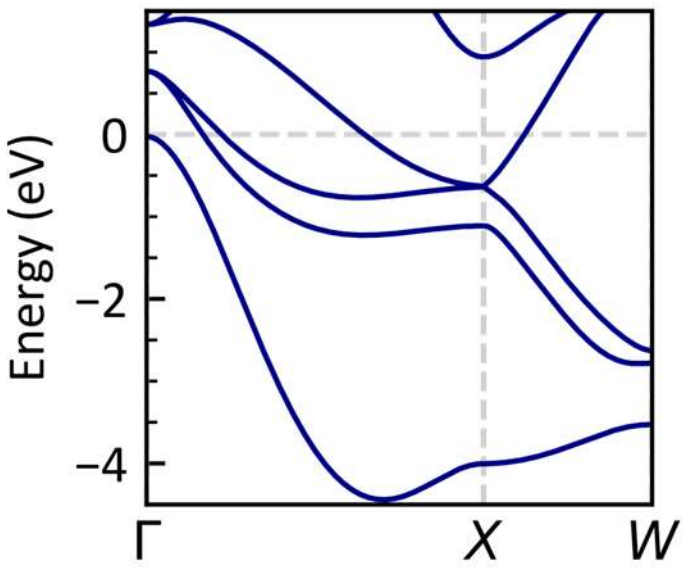

**Figure S1.** Wide energy range *E-k* dispersion based on the constructed TB model for unstrained GdSb.

**Table S1.** Nearest- and next-nearest neighbors TB parameters for GdSb, extracted from DFT calculations.

| Parameters (eV)/Strain | −2% | 0% | +2% |
|---|---|---|---|
| $p_{Sb}p_{Sb}\sigma$ | 0.713 | 0.661 | 0.583 |
| $p_{Sb}p_{Sb}\pi$ | -0.095 | -0.090 | -0.064 |
| $d_{Gd}d_{Gd}\sigma$ | -0.814 | -0.745 | -0.695 |
| $d_{Gd}d_{Gd}\pi$ | 0.233 | 0.239 | 0.238 |
| $d_{Gd}d_{Gd}\delta$ | 0.051 | 0.046 | 0.056 |
| $p_{Sb}d_{Gd}\pi$ | -0.964 | -0.894 | -0.845 |
| $p_{Sb}d_{Gd}\sigma$ | -1.899 | -1.745 | -1.665 |
| $\lambda_{Gd}$ | | 0.43 | |
| $\lambda_{Sb}$ | | 0.53 | |

## III. DFT calculations

The Gd *4f* electrons were treated as valence electrons for antiferromagnet (AFM) calculations in Fig. 2(g-h), whereas for the nonmagnetic phase calculation in Figure S2 the *4f* electrons were treated as core electrons. The configurations of the valence shells of Gd is $4f^7 5s^2 5p^6 5d^2 6s^1$ (with *4f* electrons treated as core electrons) and Sb is $5s^2 5p^3$. We used the calculated equilibrium lattice parameter of 6.197 Å for GdSb to consistently determine the Poisson ratio. Kohn-Sham orbitals in DFT[6,7] were expanded using a plane-wave basis set with the value of energy cutoff of 400 eV. Interactions between ion cores and valence electrons were described by the projector augmented wave (PAW) method[8]. We used a rhombohedral unit cell consisting of 4 atoms for magnetic phase calculations to simulate the AFM state in GdSb. The local magnetic moment on each Gd site was constrained to point along $[11\bar{2}]$ direction. An $8 \times 8 \times 8$ $\Gamma$-centered k-point mesh was used for integration over the first Brillouin zone. The folded band structure of the AFM unit cell was then unfolded back to the FCC primitive unit cell to facilitate a direct comparison with ARPES data[9,10].

The Fermi surface and Fermi volume were obtained by Wannier interpolation from first principles using the nonmagnetic phase. Since the Gd *d* and Sb *p* are relevant orbitals around the Fermi level, they were used as the starting projectors for the Wannier orbitals construction. The charge carrier concentrations were then estimated from the obtained Fermi volumes by using the SKEAF code[11]. Hopping term values used in constructing the TB model were extracted from the Wannier orbitals-based Hamiltonian, which is directly achieved from the wannierization process.

We have further investigated the consequence of the size quantization effect on the 4 nm thin films by performing calculations for 4-nm-thick freestanding slabs in Figure S2. Unstrained and in-plane strained cases were simulated using a supercell consisting of 13 monolayers (ML, 2 ML in one unit cell) thick slabs along the [001] direction, assuming a nonmagnetic phase. A vacuum space of at least 20 Å was included in the [001] direction of the supercell to remove artificial interaction between images of the slabs.

## IV. ARPES- and DFT-extracted Fermi wave vectors and band extrema

Table 1 lists band positions collected via ARPES and results from DFT calculations. The values of hole and electron carrier Fermi wave vectors for both GdSb strain levels are similar to other bulk crystal RE-V compounds studied with ARPES[12–17]. Comparison to other RE-Vs also demonstrates that for in-plane dispersions, biaxial lattice compression shows a similar trend as chemical pressure induced by lanthanide contraction; a smaller lattice parameter leads to a higher chemical potential[17,18]. The experimental Fermi velocity of both the hole and electron carriers does not show a significant change as a function of strain. This can be explained by the relatively high energy separation of the Fermi level from the band extrema, where most effective mass changes are expected to take place[19].

The hole band Fermi wave vectors and γ band maximum in Table 1 (extracted from Fig. 2(f), Fig. 3, and Fig. 4) deviate from our DFT calculations, which predicted a displacement of the valence-band maximum with strain (Fig. 2(g-h) and Table 1), a trend we do not observe experimentally. In Fig. 4, the hole split-off γ band maximum at Γ does not shift significantly with strain and remains at the same maximum energy of −0.3 eV. In addition, when transitioning from tensile to compressive strain we observe a decrease in $k_F$ along $\bar{M} - \bar{\Gamma} - \bar{M}$ in contrast to DFT predictions. These two trends suggest that the observed Fermi level position in the compressive film could differ from the initial DFT-calculated position shown in Fig. 2(g-h). The origin of this variation between ARPES results and DFT calculations could be either due to quantum size effects in the hole band[1], related to the hydrostatic tensor contributions to the DFT-modeled valence band shift[20], or resulting from experimental effects such as defects in the compressively strained film leading to bulk doping or potential Fermi level pinning at the film surface.

## V. Confinement effects

In Figure S2, the 4 nm strained GdSb films showed an experimental shift in the γ hole band by ~0.1 eV away from the Fermi level and reduced Fermi wave vectors in both the light- and heavy-hole bands along $\bar{X} - \bar{\Gamma} - \bar{X}$ compared to the bulk-limit 20-nm-thick unstrained film[21]. The electron pocket second quantum well state at $\bar{M}$ is predicted to nearly graze the Fermi level in our calculations in Figure S2 but experimentally lies well below the Fermi level in Fig. 2(b). The smaller spacing of quantum well states is most likely due to the finite potential barrier at the film-substrate/buffer layer interface, whereas our calculations assume a freestanding GdSb slab.

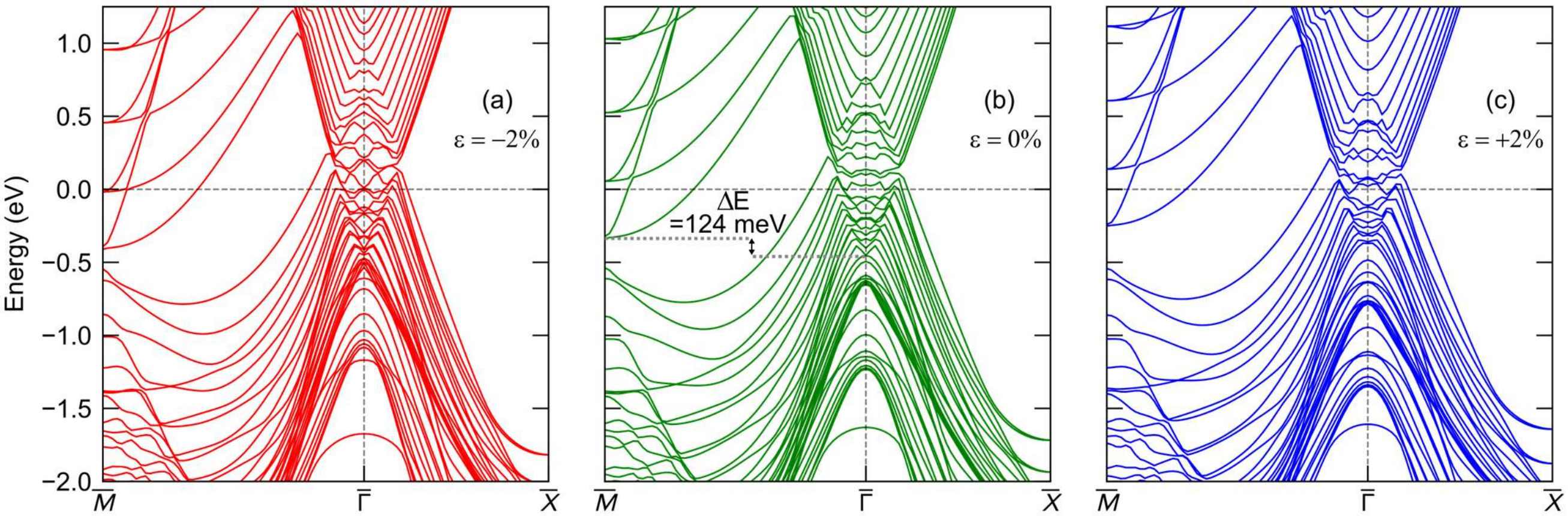

**Figure S2.** DFT-calculated electronic structure of free-standing 13-ML-thick GdSb (001) films for (a) 2% compressively strained, (b) unstrained, and (c) 2% tensile strained films. The electron pockets lying in the film plane ($X_{1,2}$, projecting to $\overline{M}$) shift to higher energies compared to the electron pockets in the film plane normal direction (Z, projecting to $\overline{\Gamma}$), with an energy difference of 124 meV calculated for the unstrained film.

## VI. Magnetic properties of strained GdSb films

GdSb can be considered an orbitally quenched system since the $4f^7$ configuration results in zero orbital angular momentum. Therefore, there is a negligible crystalline electric field (CEF) effect. The weak CEF effect in GdSb permits us to overlook the impact of strain on local symmetry breaking and to easily elucidate the role strain-driven band structure modifications and orbital overlap might have on magnetic ordering in GdSb. The $4f$ electrons in GdSb are well localized, with occupied states at 8.7 eV below the Fermi level[25]. Superexchange interactions (via *p-d* hopping in Gd-Sb) and Ruderman-Kittel-Kasuya-Yosida (RKKY) indirect-exchange interactions are expected to coexist in this compound[26], the former leading to antiferromagnetic behavior, and the latter contributing to a competing ferromagnetic order. Assuming a molecular field approximation can describe well the ordering in GdSb[27], the Néel Temperature ($T_N$) of a type II Heisenberg AFM with S=7/2 is[28]:

$$k_B T_N = \frac{2}{3} S(S+1)\hbar^2(-6J_2) = -63\hbar^2 J_2$$

$J_2$ being the next-nearest neighbor exchange constant. The pnictogen *p*-orbitals mediate the AFM superexchange interactions between the Gd atoms, and $J_2$ can be expressed by the empirical relation used for transition-metal compounds: $J_2^{super} = -\frac{n_d t_{pd}^4}{\Delta^2}\left(\frac{1}{U}+\frac{1}{\Delta}\right)$ where $n_d$ is the *d* moment induced by intra-atomic *5f-4d* exchange, $t_{pd}$ is the hopping integral between the Sb-*p* and Gd-*d* orbitals, $U$ is the on-site coulomb energy and $\Delta$ is the energy difference between the *d* and *p* orbitals [26,29]. From Table S1, we can see that both $t_{pd\sigma}$ and $t_{pd\pi}$ hopping terms increase (in absolute value) moving from tensile to compressive strain, in agreement with the stronger superexchange interaction expected with decreasing lattice constant. A monotonic increase in $T_N$ (including the lattice-matched film results[21]) is measured as a function of strain. Figure S3 highlights the change in $T_N$ as a function of film strain with a 2.6 ºC difference measured. Similar strain-dependent changes were measured for EuTe, an S=7/2 rare-earth analog[30]. However, the superexchange parameter $J_2$ in GdSb is predicted to be much larger and more sensitive to variations in the lattice constant[27].

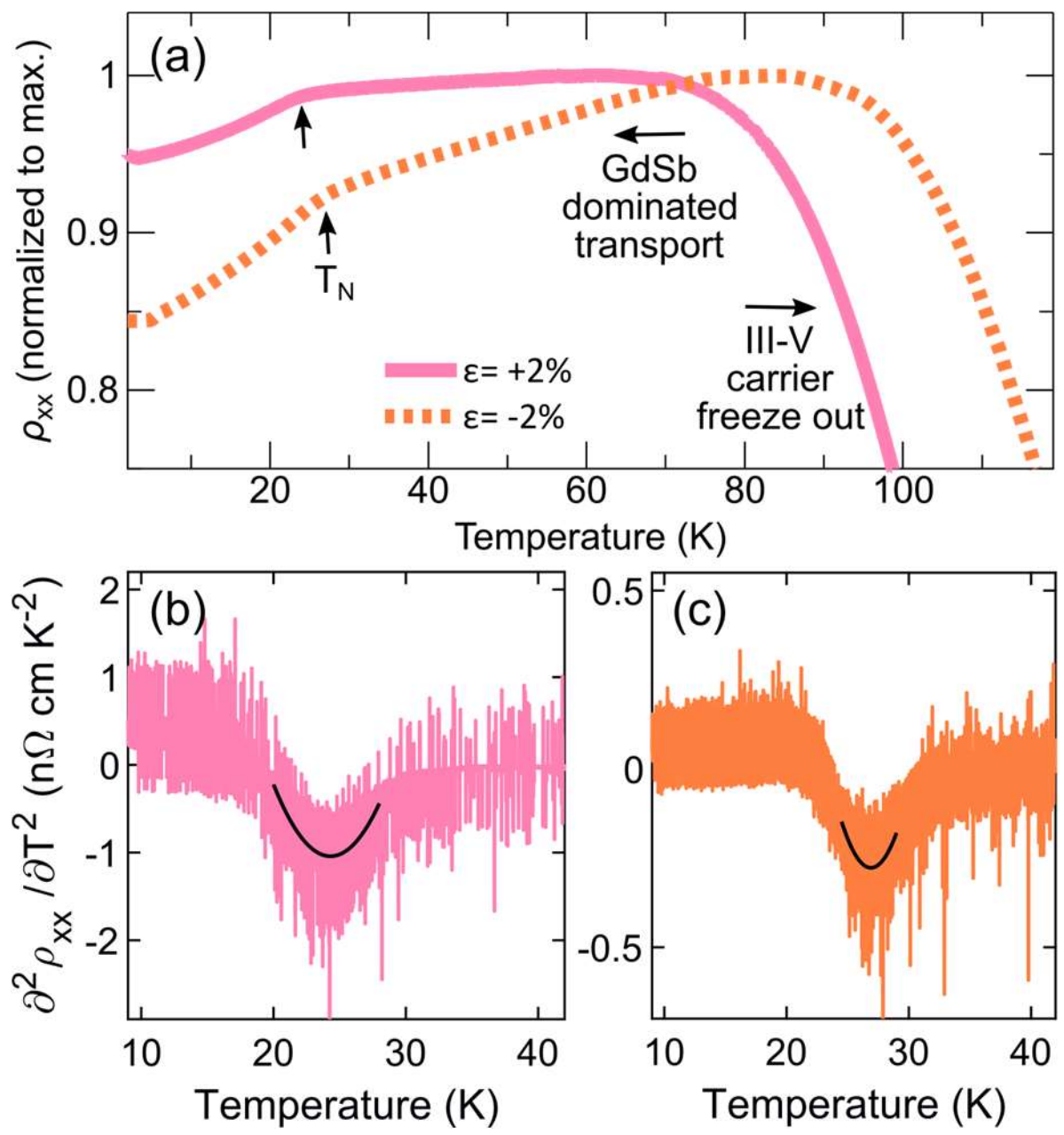

**Figure S3.** Néel temperature in strained 4-nm-thick GdSb films. (a) Temperature dependence of longitudinal resistivity, normalized to the maximum resistivity. At high temperatures > 70 K, transport is dominated by the III-V substrate and/or buffer layer. Under 70 K, the charge carriers in the III-V buffer layer and GaSb substrate freeze out, and transport is dominated by the GdSb films. (b-c) Néel temperature extracted from a parabolic fit (black line) to the second derivative of the resistivity at B=0 T, for (b) +2% strain ($T_N$=24.3 K) and (c) −2% biaxial strain ($T_N$=26.9 K).